\input amstex
\documentstyle{amsppt}
%
\catcode`@=11
\redefine\output@{%
  \def\break{\penalty-\@M}\let\par\endgraf
  \ifodd\pageno\global\hoffset=105pt\else\global\hoffset=8pt\fi  
  \shipout\vbox{%
    \ifplain@
      \let\makeheadline\relax \let\makefootline\relax
    \else
      \iffirstpage@ \global\firstpage@false
        \let\rightheadline\frheadline
        \let\leftheadline\flheadline
      \else
        \ifrunheads@ 
        \else \let\makeheadline\relax
        \fi
      \fi
    \fi
    \makeheadline \pagebody \makefootline}%
  \advancepageno \ifnum\outputpenalty>-\@MM\else\dosupereject\fi
}
\catcode`\@=\active
\nopagenumbers
\def\negskp{\hskip -2pt}
\accentedsymbol\vnabla{\nabla\kern -7pt\raise 5pt\vbox{\hrule width 5.3pt}
\kern 1.7pt}
\accentedsymbol\brbG{\overset\hbox{\kern -1.5pt\vrule depth 1pt 
 height 0.4pt\vbox{\hrule width 4.5pt}\vrule depth 1pt 
 height 0.4pt}\to{\bold G}}
\accentedsymbol\brG{\overset\hbox{\kern 0.5pt\vrule depth 1pt 
 height 0.4pt\vbox{\hrule width 4pt}\vrule depth 1pt 
 height 0.4pt}\to G}
\accentedsymbol\brbR{\overset\hbox{\kern -1.5pt\vrule depth 1pt 
 height 0.4pt\vbox{\hrule width 4pt}\vrule depth 1pt 
 height 0.4pt}\to{\bold R}}
\accentedsymbol\brR{\overset\hbox{\kern 0.5pt\vrule depth 1pt 
 height 0.4pt\vbox{\hrule width 4pt}\vrule depth 1pt 
 height 0.4pt}\to R}
\def\blue#1{#1}
\catcode`#=11\def\diez{#}\catcode`#=6
\def\mycite#1{\cite{\blue{#1}}\immediate\special{ps:
     ShrHPSdict begin /ShrBORDERthickness 0 def}}
\def\myciterange#1#2{\cite{\blue{#2}}\immediate\special{ps:
     ShrHPSdict begin /ShrBORDERthickness 0 def}}
\def\mytag#1{%
    \tag#1}
\def\mythetag#1{\thetag{\blue{#1}}\immediate\special{ps:
     ShrHPSdict begin /ShrBORDERthickness 0 def}}
\def\myrefno#1{\no#1}
\def\myhref#1#2{\blue{#2}\immediate\special{ps:
     ShrHPSdict begin /ShrBORDERthickness 0 def}}
\def\myEarXivlink{\myhref{http://arXiv.org}{http:/\negskp/arXiv.org}}
\def\myFreeTextbooks{\myhref{http://www.freetextbooks.boom.ru}
{http:/\negskp/freetextbooks.boom.ru}}
\def\mytheorem#1{\csname proclaim\endcsname{Theorem #1}}

\def\myconjecture#1{\csname proclaim\endcsname{Conjecture #1}}
\def\mytheconjecture#1{\blue{#1}\immediate\special{ps:
     ShrHPSdict begin /ShrBORDERthickness 0 def}}
\pagewidth{360pt}
\pageheight{606pt}
\topmatter
\title
A note on the dynamics and thermodynamics of dislocated crystals.
\endtitle
\author
Ruslan Sharipov
\endauthor
\address Rabochaya street 5, 450003 Ufa, Russia\newline
\vphantom{.}\kern 12pt Phone: +7\,(917)\,476\ 9348
\endaddress
\urladdr
\myhref{http://www.geocities.com/r-sharipov}
{http:/\negskp/www.geocities.com/r-sharipov}
\endurladdr
\abstract
    The dynamics and thermodynamics of dislocated crystals 
are studied within the framework of the nonlinear theory of 
elastic and plastic deformations.
\endabstract
\endtopmatter
\loadbold
\TagsOnRight
\document

\rightheadtext{A note on the dynamics and thermodynamics \dots}
\head
1. Introduction.
\endhead
    In the series of papers \myciterange{1}{1--4} the kinematics of
crystalline media in the presence of dislocations was studied. Here
we consider the dynamics and thermodynamics of a crystalline medium.
This paper continues the series of papers \myciterange{1}{1--4}.
Therefore, we do not provide a large introductory section in it. For 
more details the reader is referred to the previous papers.
\head
2. Specific free energy function.
\endhead
    By analogy to \mycite{5}, we shall construct the theory on the
base of the specific free energy function $f$ (the amount of free 
energy per unit mass of a medium). The deformation state of a 
crystalline medium in the presence of dislocations is described
by the {\it incompatible distorsion tensor\/} $\hat\bold T$. This 
is a {\it double space tensor\/} with one upper index\footnote{ \, 
It is implicitly assumed that some coordinate systems in $\Bbb B$ 
and $\Bbb E$ are chosen and $\hat\bold T$ is represented by its 
components $\hat T^i_j$ respective to those coordinate systems.}\ 
associated with the {\it Burgers space\/} $\Bbb B$  and with one lower 
index associated with the {\it real space\/} $\Bbb E$. The other physically
important parameter is the tensor of the {\it Burgers vector density\/}
$\boldsymbol\rho$. It is also a double space tensor with two indices.
Like in the case of $\hat\bold T$, the upper index of $\boldsymbol\rho$
is associated with the Burgers space, while its lower index is
associated with the real space. The easiest way of choosing the specific
free energy function $f$ is to write it as
\adjustfootnotemark{-1}
$$
\hskip -2em
f=f(T,\hat\bold T,\boldsymbol\rho),
\mytag{2.1}
$$
where $T$ is the temperature. However, we shall not write $f$ in 
this simplest way \mythetag{2.1}. The matter is that $\hat\bold T$ 
and $\boldsymbol\rho$ contain some excessive amount of purely geometric
information absolutely inessential for describing the thermodynamics of 
a medium. We need to make some efforts to get rid of this inessential 
part of information in the tensors $\hat\bold T$ and $\boldsymbol\rho$.
\par
     The Burgers space $\Bbb B$, as it was defined in \mycite{1}, is a 
copy of the real space $\Bbb E$. It is assumed to be filled with the 
ideal \pagebreak (dislocation-free and non-distorted) crystalline medium
(see Fig\.~2.2) identical to that we consider in the real space. The 
tensor field $\hat\bold T$ determines a linear mapping at each point $p$ 
of the real space: 
$$
\hskip -2em
\hat\bold T\!:\,T_{\!p\,}\Bbb E\to\Bbb B.
\mytag{2.2}
$$
Here $T_{\!p\,}\Bbb E$ is the set of vectors attached to the point 
$p\in\Bbb E$. In geometry, when $\Bbb E$ is treated as a manifold, 
such a space is called a {\it tangent space}. Both spaces $\Bbb E$ 
and $\Bbb B$ are equipped with the Euclidean metrics $\bold g$ and
$\overset\sssize\star\to{\bold g}$ respectively. The difference is
that the Burgers space $\Bbb B$ in \mythetag{2.2} is treated as a 
\vadjust{\vskip 5pt\hbox to 0pt{\kern -30pt
\includegraphics{crdyn01.eps}\hss}\vskip 180pt}linear 
space, while $\Bbb E$ is treated as a point space (or a manifold). 
The mapping \mythetag{2.2} is non-degenerate: $\det\hat\bold T\neq 0$.
Hence, one can consider the inverse mapping
$$
\hskip -2em
\hat\bold S\!:\,\Bbb B\to T_{\!p\,}\Bbb E.
\mytag{2.3}
$$
The inverse mapping \mythetag{2.3} is given by the {\it inverse
incompatible distorsion tensor\/} $\hat\bold S$. This is also a 
double space tensor with two indices: its upper index is associated 
with $\Bbb E$ and its lower index is associated with $\Bbb B$.\par
     Suppose that the point $p$ is fixed. The rotation of the crystalline
body as a whole about the point $p\in\Bbb E$ does not change its physical
state (see Fig\.~2.1). Such a rotation is expressed by the following 
transformations at the point $p$:
$$
\xalignat 2
&\hskip -2em
\hat T^i_k\to\sum^3_{\alpha=1}O^{\,\alpha}_k\ \hat T^i_\alpha,
&&\rho^i_k\to\sum^3_{\alpha=1}O^{\,\alpha}_k\ \rho^i_\alpha.
\mytag{2.4}
\endxalignat
$$
Here $O^{\,\alpha}_k$ are the components of some rotation 
matrix\footnote{ \ Note that the transformations \mythetag{2.4} are
different from those in section~2 of \mycite{3} and from those in 
formulas \thetag{3.19}, \thetag{3.20}, \thetag{3.21} in \mycite{4}.}.
Since the physical state of a crystalline body is invariant under the
transformations \mythetag{2.4}, so should be the free energy function
\mythetag{2.1}. Let's define the following tensors $\brbG$ and $\brbR$:
$$
\xalignat 2
&\hskip -2em
\brG_{ij}=\sum^3_{\alpha=1}\sum^3_{\beta=1}g_{\alpha\beta}\ 
\hat S^\alpha_i\ \hat S^\beta_j,
&&\brR^{\,i}_j=\sum^3_{\alpha=1}\rho^i_\alpha\ \hat S^\alpha_j.
\mytag{2.5}
\endxalignat
$$
Both indices $i$ and $j$ in \mythetag{2.5} are associated with the
Burgers space $\Bbb B$. Nevertheless, $\brbG$ and $\brbR$ are double
space tensors because their components are functions of the coordinates
of a point $p\in\Bbb E$ and of the time variable $t$:
$$
\xalignat 2
&\hskip -2em
\brG_{ij}=\brG_{ij}(t,y^1,y^2,y^3),
&&\brR^{\,i}_j=\brR^{\,i}_j(t,y^1,y^2,y^3).
\mytag{2.6}
\endxalignat
$$
Note that the tensors $\brbG$ and $\brbR$ are different from the purely
real space tensors $\hat\bold G$ and $\bold R$ introduced in \mycite{1}
and in \mycite{3}. Their components are given by the formulas
$$
\xalignat 2
&\hskip -2em
\hat G_{\alpha\beta}=\sum^3_{i=1}\sum^3_{j=1}
\overset\sssize\star\to g_{ij}\ 
\hat T^i_\alpha\ \hat T^j_\beta,
&&R^{\,\alpha}_\beta=\sum^3_{i=1}\hat S^\alpha_i\ \rho^i_\beta.
\mytag{2.7}
\endxalignat
$$
The tensor $\hat\bold G$ given by the first formula \mythetag{2.7} is 
called the {\it elastic deformation tensor}. As for $\bold R$, it is 
the purely real space representation of the Burgers vector density
tensor $\boldsymbol\rho$.\par
    Now let's return back to the tensors $\brbG$ and $\brbR$ in
\mythetag{2.5} and find that they are invariant with respect to the 
transformations \mythetag{2.4}. Therefore, we can use $\brbG$ and 
$\brbR$ as independent variables instead of $\bold T$ and 
$\boldsymbol\rho$ in \mythetag{2.1}:
$$
\hskip -2em
f=F(T,\brbG,\overset\sssize\star\to{\bold g},\brbR,\rho).
\mytag{2.8}
$$
Note that the function $F$ has no explicit dependence on the coordinates
of a point $p\in\Bbb E$. This function represents the refined properties 
of a crystal in its homogeneous state (see Fig\.~2.2). It should be
tabulated on the base of experiments. Note also that, apart from $\brbG$ 
and $\brbR$, we introduced the additional arguments $\overset\sssize\star
\to{\bold g}$ and $\rho$ in \mythetag{2.8}, where $\rho$ is the density
of a crystal and $\overset\sssize\star\to{\bold g}$ is the metric tensor 
of the Euclidean metric in the Burgers space $\Bbb B$. The constant tensor
$\overset\sssize\star\to{\bold g}$ does not change the time and coordinate
dependence of $f$ determined by \mythetag{2.6} and by the scalar functions
$$
\xalignat 2
&T=T(t,y^1,y^2,y^3),
&&\rho=\rho(t,y^1,y^2,y^3).
\endxalignat
$$\par
    Having written \mythetag{2.8} in place of \mythetag{2.1}, we have
reached one of our goals --- we have got the rotationally invariant free
energy function. The experimental data should go to the theory through
the function $F$. However, for the sake of beauty, in the formulas
below we need to express $f$ through some purely real space tensor fields.
The tensor $\brbR$ is expressed through $\bold R$ as follows:
$$
\hskip -2em
\brR^{\,i}_j=\sum^3_{\alpha=1}\sum^3_{\beta=1}\hat T^i_\beta\ 
\hat S^\alpha_j\ R^\beta_\alpha.
\mytag{2.9}
$$
The tensor $\brbG$ cannot be expressed through $\hat\bold G$, but it is  
expressed through the metric tensor $\bold g$ of the real space $\Bbb E$
according to the first formula \mythetag{2.5}:
$$
\hskip -2em
\brG_{ij}=\sum^3_{\alpha=1}\sum^3_{\beta=1}g_{\alpha\beta}\ 
\hat S^\alpha_i\ \hat S^\beta_j.
\mytag{2.10}
$$
Similarly, the metric tensor $\overset\sssize\star\to{\bold g}$ in
\mythetag{2.8} is expressed through $\hat\bold G$ as follows:
$$
\hskip -2em
\overset\sssize\star\to g_{ij}
=\sum^3_{\alpha=1}\sum^3_{\beta=1}\hat G_{\alpha\beta}\ 
\hat S^\alpha_i\ \hat S^\beta_j.
\mytag{2.11}
$$
This formula \mythetag{2.11} is inverse to the first formula 
\mythetag{2.7}. Substituting \mythetag{2.9}, \mythetag{2.10}, and
\mythetag{2.11} into \mythetag{2.8}, we can write \mythetag{2.8} 
in the following form:
$$
\hskip -2em
f=f(t,p,T,\bold g,\hat\bold G,\bold R,\rho).
\mytag{2.12}
$$
We do not keep $\hat\bold S$ and $\hat\bold T$ in \mythetag{2.12}
as independent variables since they are not purely real space 
tensor fields. Their presence is reflected by the arguments $t$
and $p$, where $t$ is the time variable and $p$ is a point of the
real space $\Bbb E$. In a coordinate presentation of $f$ its 
argument $p$ is replaced by the variables $y^1,\,y^2,\,y^3$.
\head
3. Specific free energy function\\
as an extended scalar field.
\endhead
     The specific free energy function \mythetag{2.12} is a scalar
function with three tensorial arguments $\bold g$, $\hat\bold G$,
and $\bold R$. In this form it fits the definition of an extended
scalar field (see definition~4.1 in \mycite{6}). The theory of extended
tensor fields was especially derived in \mycite{6} for to use it in the
present paper. The extended scalar field $f$ in \mythetag{2.12} is not
an arbitrary extended scalar field. It is produced from the function
$F$ in \mythetag{2.8}. For this reason it satisfies some partial 
differential equations written in terms of covariant and multivariate
derivatives defined in \mycite{6}.
\mytheorem{3.1} The specific free energy function $f$ of a dislocated
crystalline medi\-um is an extended scalar field satisfying the differential
equation
$$
\hskip -2em
\gathered
\nabla_{\!\gamma}f=
-\sum^3_{m=1}\sum^3_{\alpha=1}\sum^3_{\beta=1}
\vnabla^{\alpha\beta}[2]f
\left(\hat Z^m_{\gamma\alpha}\ g_{m\beta}
+\hat Z^m_{\gamma\beta}\ g_{\alpha m}\right)-\\
-\sum^3_{m=1}\sum^3_{\alpha=1}\sum^3_{\beta=1}
\vnabla^{\alpha\beta}[3]f\left(\hat Z^m_{\gamma\alpha}
\ \hat G_{m\beta}+\hat Z^m_{\gamma\beta}
\ \hat G_{\alpha m}\right)+\\
+\sum^3_{n=1}\sum^3_{m=1}\sum^3_{\alpha=1}\sum^3_{\beta=1}
\vnabla^\alpha_{\!\beta}[4]f\left(\hat Z^\beta_{\gamma m}
\ R^m_\alpha-\hat Z^m_{\gamma\alpha}\ R^\beta_m\right)\!,
\endgathered
\mytag{3.1}
$$
where $\hat Z^k_{ij}$ are the components of the tensor field 
$\hat\bold Z$ introduced in paper \mycite{3}:
$$
\hskip -2em
\hat Z^k_{ij}=\sum^3_{m=1}\hat S^k_m\,\nabla_i\,\hat T^m_j
=\sum^3_{m=1}\hat S^k_m\left(\frac{\partial\hat T^m_j}{\partial y^i}
-\shave{\sum^3_{n=1}}\Gamma^n_{ij}\ \hat T^m_n\right)\!.
\mytag{3.2}
$$
\endproclaim
\demo{Proof} First of all let's calculate the multivariate derivatives
$\vnabla^{\alpha\beta}[2]f$, $\vnabla^{\alpha\beta}[3]f$,\linebreak and
$\vnabla^\alpha_{\!\beta}[4]f$ defined due to three tensorial arguments
in \mythetag{2.12}:
$$
\allowdisplaybreaks
\align
&\hskip -2em
\vnabla^{\alpha\beta}[2]f=\frac{\partial f}{\partial g_{\alpha\beta}}
=\sum^3_{i=1}\sum^3_{j=1}\frac{\partial F}{\partial\brG_{ij}}
\,\hat S^\alpha_i\,\hat S^\beta_j,
\mytag{3.3}\\
&\hskip -2em
\vnabla^{\alpha\beta}[3]f=\frac{\partial f}{\partial\hat G_{\alpha\beta}
\vphantom{\vrule height 10pt}}=\sum^3_{i=1}\sum^3_{j=1}\frac{\partial F}
{\partial\overset\sssize\star\to g_{ij}}\,\hat S^\alpha_i\,\hat S^\beta_j,
\mytag{3.4}\\
&\hskip -2em
\vnabla^\alpha_{\!\beta}[4]f=\frac{\partial f}{\partial R^\beta_\alpha
\vphantom{\vrule height 11pt}}=\sum^3_{i=1}\sum^3_{j=1}\frac{\partial F}
{\partial\brR^{\,i}_j}\,\hat S^\alpha_j\,\hat T^i_\beta.
\mytag{3.5}
\endalign
$$
Then we calculate the covariant derivatives of the extended scalar field
\mythetag{2.12} applying the formula \thetag{13.15} from \mycite{6}:
$$
\hskip -2em
\gathered
\nabla_{\!\gamma}f=\frac{\partial f}{\partial y^\gamma}
+\sum^3_{m=1}\sum^3_{\alpha=1}\sum^3_{\beta=1}
\left(\Gamma^m_{\gamma\alpha}\,g_{m\beta}+\Gamma^m_{\gamma\beta}
\,g_{\alpha m}\right)\frac{\partial f}{\partial g_{\alpha\beta}}\,+\\
+\sum^3_{m=1}\sum^3_{\alpha=1}\sum^3_{\beta=1}
\left(\Gamma^m_{\gamma\alpha}\,\hat G_{m\beta}+\Gamma^m_{\gamma\beta}
\,\hat G_{\alpha m}\right)\frac{\partial f}
{\partial\hat G_{\alpha\beta}\vphantom{\vrule height 10pt}}\,+\\
+\sum^3_{m=1}\sum^3_{\alpha=1}\sum^3_{\beta=1}\left(\Gamma^m_{\gamma \alpha}
\,R^\beta_m-\Gamma^\beta_{\gamma m}\,R^m_\alpha\right)
\frac{\partial f}{\partial R^\beta_\alpha}.
\endgathered
\mytag{3.6}
$$
Writing \mythetag{3.6}, we assume that $y^1,\,y^2,\,y^3$ are some
curvilinear coordinates in the real space $\Bbb E$ and $\Gamma^k_{ij}$
are the components of the metric connection for the Euclidean metric
$\bold g$ in $\Bbb E$. Now let's calculate the partial derivatives
$\partial f/\partial y^\gamma$:
$$
\gather
\frac{\partial f}{\partial y^\gamma}
=\sum^3_{i=1}\sum^3_{j=1}\sum^3_{\alpha=1}\sum^3_{\beta=1}
\frac{\partial F}{\partial\brG_{ij}}\ g_{\alpha\beta} 
\left(\frac{\partial\hat S^\alpha_i}{\partial y^\gamma}
\ \hat S^\beta_j+\hat S^\alpha_i\ 
\frac{\partial\hat S^\beta_j}{\partial y^\gamma}\right)+\\
+\sum^3_{i=1}\sum^3_{j=1}\sum^3_{\alpha=1}\sum^3_{\beta=1}
\frac{\partial F}{\partial\overset\sssize\star\to g_{ij}}
\ \hat G_{\alpha\beta}\left(\frac{\partial\hat S^\alpha_i}
{\partial y^\gamma}\ \hat S^\beta_j+\hat S^\alpha_i\ 
\frac{\partial\hat S^\beta_j}{\partial y^\gamma}\right)+\\
+\sum^3_{i=1}\sum^3_{j=1}\sum^3_{\alpha=1}\sum^3_{\beta=1}
\frac{\partial F}{\partial\brR^{\,i}_j}\ R^\beta_\alpha
\left(\frac{\partial\hat T^i_\beta}{\partial y^\gamma}
\ \hat S^\alpha_j+\hat T^i_\beta\ \frac{\partial\hat S^\alpha_j}
{\partial y^\gamma}\right).
\endgather
$$
The matrices $\hat\bold S$ and $\hat\bold T$ are inverse to 
each other. Applying this fact and the equalities \mythetag{3.3}, 
\mythetag{3.4}, \mythetag{3.5} to the above formula, we obtain
$$
\gather
\frac{\partial f}{\partial y^\gamma}=
-\sum^3_{n=1}\sum^3_{m=1}\sum^3_{\alpha=1}\sum^3_{\beta=1}
\frac{\partial f}{\partial g_{\alpha\beta}}\left(\hat S^m_n\,
\frac{\partial \hat T^n_\alpha}{\partial y^\gamma}\ g_{m\beta}
+\hat S^m_n\,\frac{\partial \hat T^n_\beta}{\partial y^\gamma}
\ g_{\alpha m}\right)-\\
-\sum^3_{n=1}\sum^3_{m=1}\sum^3_{\alpha=1}\sum^3_{\beta=1}
\frac{\partial f}{\partial\hat G_{\alpha\beta}
\vphantom{\vrule height 10pt}}\left(\hat S^m_n\,
\frac{\partial \hat T^n_\alpha}{\partial y^\gamma}\ \hat G_{m\beta}
+\hat S^m_n\,\frac{\partial \hat T^n_\beta}{\partial y^\gamma}
\ \hat G_{\alpha m}\right)+\\
+\sum^3_{n=1}\sum^3_{m=1}\sum^3_{\alpha=1}\sum^3_{\beta=1}
\frac{\partial f}{\partial R^\beta_\alpha
\vphantom{\vrule height 11pt}}\left(\hat S^\beta_n
\,\frac{\partial \hat T^n_m}{\partial y^\gamma}\ R^m_\alpha
-\hat S^m_n\,\frac{\partial \hat T^n_\alpha}{\partial y^\gamma}
\ R^\beta_m\right)\!.
\endgather
$$
Now, substituting this expression for $\partial f/\partial y^\gamma$
into \mythetag{3.6} and taking into account \mythetag{3.2}, we derive
the required equality \mythetag{3.1}. The theorem is proved.
\qed\enddemo
     Note that $\Gamma^k_{ij}$ in \mythetag{3.2} and \mythetag{3.6}
are the components of the symmetric Euclidean connection $\Gamma$ associated
with the standard Euclidean metric $\bold g$ in the real space $\Bbb E$.
These quantities are equal to zero in Cartesian coordinates and arise 
only if one chooses curvilinear coordinates $y^1,\,y^2,\,y^3$. Let's
remember that there is another connection $\hat\Gamma$ in the theory.
Its components are given by the formula \thetag{2.3} in \cite{4}:
$$
\hskip -2em
\hat\Gamma^k_{ij}=\Gamma^k_{ij}+\hat Z^k_{ij}.
\mytag{3.7}
$$
This connection \mythetag{3.7} is associated with the tensor field
$\hat\bold G$ treated as a metric. If we replace $\Gamma$ by $\hat
\Gamma$ in \mythetag{3.6}, then the equation \mythetag{3.1} reduces
to the following one:
$$
\hskip -2em
\hat\nabla_{\!\gamma}f=0.
\mytag{3.8}
$$
\subhead Remark\endsubhead Note that both $\nabla_{\!\gamma}$ and
$\hat\nabla_{\!\gamma}$ in \mythetag{3.1} and \mythetag{3.8} are not
standard covariant derivatives. They are so-called {\it spacial 
covariant derivatives} introduced in \mycite{6}.\par
    Now let's calculate the time derivative $\partial f/\partial t$
considering $f$ as an extended scalar field \mythetag{2.12} and thus
taking $T$, $\bold g$, $\hat\bold G$, $\bold R$, and $\rho$ for 
independent variables:
$$
\gather
\frac{\partial f}{\partial t}=
-\sum^3_{n=1}\sum^3_{m=1}\sum^3_{\alpha=1}\sum^3_{\beta=1}
\frac{\partial f}{\partial g_{\alpha\beta}}\left(\hat S^m_n\,
\frac{\partial \hat T^n_\alpha}{\partial t}\ g_{m\beta}
+\hat S^m_n\,\frac{\partial \hat T^n_\beta}{\partial t}
\ g_{\alpha m}\right)-\\
-\sum^3_{n=1}\sum^3_{m=1}\sum^3_{\alpha=1}\sum^3_{\beta=1}
\frac{\partial f}{\partial\hat G_{\alpha\beta}
\vphantom{\vrule height 10pt}}\left(\hat S^m_n\,
\frac{\partial \hat T^n_\alpha}{\partial t}\ \hat G_{m\beta}
+\hat S^m_n\,\frac{\partial \hat T^n_\beta}{\partial t}
\ \hat G_{\alpha m}\right)+\\
+\sum^3_{n=1}\sum^3_{m=1}\sum^3_{\alpha=1}\sum^3_{\beta=1}
\frac{\partial f}{\partial R^\beta_\alpha
\vphantom{\vrule height 11pt}}\left(\hat S^\beta_n
\,\frac{\partial \hat T^n_m}{\partial t}\ R^m_\alpha
-\hat S^m_n\,\frac{\partial \hat T^n_\alpha}{\partial t}
\ R^\beta_m\right)\!.
\endgather
$$
The components of the time derivative $\partial\hat\bold T/\partial t$ 
are given by the formula
$$
\hskip -2em
\frac{\partial\hat T^n_\beta}{\partial t}=-j^{\,n}_\beta
-\sum^3_{i=1}\nabla_{\!\beta}v^i\,\hat T^n_i-\sum^3_{i=1}v^i
\,\nabla_{\!\beta}\hat T^n_i.
\mytag{3.9}
$$
The formula \mythetag{3.9} is derived from the formulas \thetag{4.4}
and \thetag{4.5} in \mycite{2}. These two formulas are valid only
under the assumption that the conjecture~4.1 in \mycite{2} is valid.
\subhead Remark\endsubhead Here we do not discuss the conjecture~4.1
from \mycite{2}. However, one should note that all results below in
this paper are obtained under the assumption that this conjecture is
valid.\par
     The quantities $j^{\,n}_\beta$ in \mythetag{3.9} represent the
double space tensor $\bold j$\,, it is called the tensor of the
{\it Burgers vector flow density}. The quantities $v^i$ are the
components of the velocity vector $\bold v$ describing the substance
flow in the medium. Multiplying \mythetag{3.9} by $S^m_n$ and summing
over the index $n$, we derive the following equality:
$$
\hskip -2em
\sum^3_{n=1}\hat S^m_n\,\frac{\partial\hat T^n_\beta}{\partial t}
=-J^{\,m}_\beta-\nabla_{\!\beta}v^m-\sum^3_{i=1}v^i\hat Z^{\,m}_{\beta\,i}.
\mytag{3.10}
$$
Here $J^{\,m}_\beta$ are the components of the tensor $\bold J$. 
\pagebreak This tensor is the purely real space version of the 
tensor $\bold j$\,. It is also called the tensor of the {\it Burgers 
vector flow density}. The components of the tensor $\bold J$ are given 
by the formula \thetag{2.9} in \mycite{3}:
$$
\hskip -2em
J^{\,m}_\beta=\sum^3_{n=1}\hat S^m_n\,j^{\,n}_\beta.
\mytag{3.11}
$$
Apart from \mythetag{3.11}, let's recall also the formula \thetag{2.10}
from \mycite{3}. It can be written as
$$
\hskip -2em
\theta^{\,m}_\beta=-J^{\,m}_\beta+\sum^3_{i=1}v^i\,
(\hat Z^m_{i\,\beta}-\hat Z^m_{\beta\,i}).
\mytag{3.12}
$$
The tensor $\boldsymbol\theta$ with the components \mythetag{3.12} is
called the tensor {\it of the rate of plastic relaxation}. Applying
\mythetag{3.12} to \mythetag{3.10}, we derive the following formula:
$$
\hskip -2em
\sum^3_{n=1}\hat S^m_n\,\frac{\partial\hat T^n_\beta}
{\partial t}=\theta^{\,m}_\beta-\nabla_{\!\beta}v^m
-\sum^3_{i=1}v^i\ \hat Z^{\,m}_{i\,\beta}.
\mytag{3.13}
$$
Now let's apply \mythetag{3.13} in order to transform the above expression
for $\partial f/\partial y^\gamma$:
$$
\gathered
\frac{\partial f}{\partial t}+\sum^3_{i=1}v^i\,\nabla_{\!i}f=
-\sum^3_{m=1}\sum^3_{\alpha=1}\sum^3_{\beta=1}
\vnabla^{\alpha\beta}[2]f\ 
\bigl(\theta^{\,m}_\alpha\ g_{m\beta}
-\nabla_{\!\alpha}v^m\ g_{m\beta}\ +\\
+\ \theta^{\,m}_\beta\ g_{\alpha m}
-\nabla_{\!\beta}v^m\ g_{\alpha m}\bigr)
-\sum^3_{m=1}\sum^3_{\alpha=1}\sum^3_{\beta=1}
\vnabla^{\alpha\beta}[3]f\ 
\bigl(\theta^{\,m}_\alpha\ \hat G_{m\beta}\ -\\
-\ \nabla_{\!\alpha}v^m\ \hat G_{m\beta}+
\theta^{\,m}_\beta\ \hat G_{\alpha m}
-\nabla_{\!\beta}v^m\ \hat G_{\alpha m}\bigr)
+\sum^3_{m=1}\sum^3_{\alpha=1}\sum^3_{\beta=1}
\vnabla^\alpha_{\!\beta}[4]f\times\\
\vspace{2ex}
\times\bigl(\theta^\beta_m\ R^m_\alpha
-\nabla_{\!m}v^\beta\ R^m_\alpha
-\theta^{\,m}_\alpha\ R^\beta_m
+\nabla_{\!\alpha}v^m\ R^\beta_m\bigr).
\endgathered\qquad
\mytag{3.14}
$$
The result obtained in the formula \mythetag{3.14} is formulated as
the following theorem.
\mytheorem{3.2}The specific free energy function $f$ of a dislocated
crystalline medi\-um is an extended scalar field satisfying the 
differential equation \mythetag{3.14}.
\endproclaim
\head
4. Balance equations. The dynamics and thermodynamics.
\endhead
     Like in \mycite{5}, here we shall develop the theory on the base 
of the balance equations. The first balance equation is very simple. 
This is the mass balance equation:
$$
\hskip -2em
\partial_t\rho
+\sum^3_{k=1}\nabla_{\!k}(\,\rho\,v^k)=0.
\mytag{4.1}
$$
The momentum balance equation looks not much more complicated than
\mythetag{4.1}:
$$
\hskip -2em
\partial_t(\,\rho\,v^i)+\sum^3_{k=1}\nabla_{\!k}\Pi^{ik}=f^i.
\mytag{4.2}
$$
However, the whole complexity of this equation is hidden in the
tensor $\boldsymbol\Pi$. As for the quantities $f^i$ in the right
hand side of \mythetag{4.2}, they are the components of the vector 
$\bold f$. They define the density of volume forces acting upon the
medium. The components of the tensor $\boldsymbol\Pi$ are given by
the following formula:
$$
\hskip -2em
\Pi^{ik}=\rho\,v^i\,v^k-\sigma^{ik}-\tilde\sigma^{ik}.
\mytag{4.3}
$$
Here $\sigma^{ik}$ and $\tilde\sigma^{ik}$ are the components of two
stress tensors $\boldsymbol\sigma$ and $\tilde{\boldsymbol\sigma}$ 
respectively. The first of them $\boldsymbol\sigma$ is the {\it regular
stress tensor}, we shall study it in more details a little bit later 
below. The second tensor $\tilde{\boldsymbol\sigma}$ in the formula 
\mythetag{4.3} is the {\it viscosity stress tensor}. Its components are
given by the formula
$$
\hskip -2em
\tilde\sigma^{ik}=\frac{1}{2}
\sum^3_{j=1}\sum^3_{q=1}\eta^{ikjq}\,
(\nabla_{\!j}v_q+\nabla_{\!q}v_j),
\mytag{4.4}
$$
where $\eta^{ikjq}$ are the components of the {\it viscosity tensor\/}
$\boldsymbol\eta$.\par
     The third balance equation expresses the {\it energy balance\/} in
a medium. This equation is written in the following form:
$$
\hskip -2em
\partial_t\!\left(\frac{\rho\,|\bold v|^2}{2}+
\rho\,\varepsilon\right)+\sum^3_{k=1}\nabla_{\!k}w^k=e.
\mytag{4.5}
$$
Here $\varepsilon$ is the {\it specific inner thermal energy}. We shall
discuss it a little bit later below along with the regular stress tensor 
$\boldsymbol\sigma$. The quantities $w^k$ in \mythetag{4.5} are the 
components of the vector $\bold w$. They are given by the formula
$$
\hskip -2em
w^k=\frac{\rho\,|\bold v|^2}{2}\,v^k+\rho\,\varepsilon\,v^k
-\sum^3_{i=1}v_i\,\sigma^{ik}-\sum^3_{i=1}v_i\,\tilde\sigma^{ik}
-\sum^3_{i=1}\nabla_{\!i}T\,\varkappa^{ik}.
\mytag{4.6}
$$
The vector $\bold w$ with the components \mythetag{4.6} is responsible
for the energy transport in a medium. It is called the {\it energy flow
density vector}. The first two terms in the right hand side of 
\mythetag{4.6} correspond to the energy transported with the mass flow.
The next two terms represent the work performed by the stress forces.
And the last term describes the {\it heat conductivity phenomenon}.
The quantities $\varkappa^{ik}$ are the components of the heat conductivity
tensor.\par
     The quantity $e$ in the right hand side of \mythetag{4.5} describes
the energy sources within the bulk of a medium. In the simplest case the
energy is produced by the work of the volume force $\bold f$. In this
case we can write
$$
\hskip -2em
e=\sum^3_{i=1}v_i\,f^i,
\mytag{4.7}
$$
where $f^i$ are the same as in \mythetag{4.2}. They are the components
of the vector $\bold f$.\par
     Our further goal is to specify the above three balance equations
\mythetag{4.1}, \mythetag{4.2}, and \mythetag{4.5} for the case of a
crystalline medium with dislocations. The specific inner thermal energy
$\varepsilon$ is related to the specific free energy $f$ as follows:
$$
\hskip -2em
\varepsilon=T\,s+f.
\mytag{4.8}
$$
Here $s$ is the {\it specific entropy}. It is derived from $f$ by means
of the formula
$$
\hskip -2em
s=-\frac{\partial f}{\partial T}.
\mytag{4.9}
$$
In this form the formula \mythetag{4.9} is applicable to \mythetag{2.8}
and to \mythetag{2.12} as well. However, applying it to \mythetag{2.12},
we can use the notations introduced in \mycite{6}:
$$
\hskip -2em
s=-\vnabla[1]f=s(t,p,T,\bold g,\hat\bold G,\bold R,\rho).
\mytag{4.10}
$$
Applying \mythetag{4.9} to the function \mythetag{2.8}, we obtain the
following equality:
$$
\hskip -2em
s=-\frac{\partial F}{\partial T}=S(T,\brbG,\overset\sssize\star
\to{\bold g},\brbR,\rho).
\mytag{4.11}
$$
The functions $S$ and $s$ in the right hand sides of  \mythetag{4.11}
and  \mythetag{4.10} are related to each other in the same way as
the functions $F$ and $f$ in  \mythetag{2.8} and  \mythetag{2.12}. For
this reason we can write the differential equations
$$
\gather
\hskip -2em
\gathered
\nabla_{\!\gamma}s=
-\sum^3_{m=1}\sum^3_{\alpha=1}\sum^3_{\beta=1}
\vnabla^{\alpha\beta}[2]s
\left(\hat Z^m_{\gamma\alpha}\ g_{m\beta}
+\hat Z^m_{\gamma\beta}\ g_{\alpha m}\right)-\\
-\sum^3_{m=1}\sum^3_{\alpha=1}\sum^3_{\beta=1}
\vnabla^{\alpha\beta}[3]s\left(\hat Z^m_{\gamma\alpha}
\ \hat G_{m\beta}+\hat Z^m_{\gamma\beta}
\ \hat G_{\alpha m}\right)+\\
+\sum^3_{n=1}\sum^3_{m=1}\sum^3_{\alpha=1}\sum^3_{\beta=1}
\vnabla^\alpha_{\!\beta}[4]s\left(\hat Z^\beta_{\gamma m}
\ R^m_\alpha-\hat Z^m_{\gamma\alpha}\ R^\beta_m\right)\!,
\endgathered
\mytag{4.12}\\
\gathered
\frac{\partial s}{\partial t}+\sum^3_{i=1}v^i\,\nabla_{\!i}s=
-\sum^3_{m=1}\sum^3_{\alpha=1}\sum^3_{\beta=1}
\vnabla^{\alpha\beta}[2]s\ 
\bigl(\theta^{\,m}_\alpha\ g_{m\beta}
-\nabla_{\!\alpha}v^m\ g_{m\beta}\ +\\
+\ \theta^{\,m}_\beta\ g_{\alpha m}
-\nabla_{\!\beta}v^m\ g_{\alpha m}\bigr)
-\sum^3_{m=1}\sum^3_{\alpha=1}\sum^3_{\beta=1}
\vnabla^{\alpha\beta}[3]s\ 
\bigl(\theta^{\,m}_\alpha\ \hat G_{m\beta}\ -\\
-\ \nabla_{\!\alpha}v^m\ \hat G_{m\beta}+
\theta^{\,m}_\beta\ \hat G_{\alpha m}
-\nabla_{\!\beta}v^m\ \hat G_{\alpha m}\bigr)
+\sum^3_{m=1}\sum^3_{\alpha=1}\sum^3_{\beta=1}
\vnabla^\alpha_{\!\beta}[4]s\times\\
\vspace{2ex}
\times\bigl(\theta^\beta_m\ R^m_\alpha
-\nabla_{\!m}v^\beta\ R^m_\alpha
-\theta^{\,m}_\alpha\ R^\beta_m
+\nabla_{\!\alpha}v^m\ R^\beta_m\bigr).
\endgathered\qquad
\mytag{4.13}
\endgather
$$
\mytheorem{4.1} The specific entropy of a dislocated crystalline medium 
$s$ is an extended scalar field satisfying the differential equations
\mythetag{4.12} and \mythetag{4.13}.
\endproclaim
It is known that $S$ is a monotonic function of $T$ in \mythetag{4.11}.
The same is true for $s$ in \mythetag{4.10}. Moreover, we have the
thermodynamical inequalities (see \mycite{7}):
$$
\xalignat 2
&\hskip -2em
\frac{\partial S}{\partial T}>0,
&&\frac{\partial s}{\partial T}>0.
\mytag{4.14}
\endxalignat
$$
This means that we can introduce the inverse functions
$$
\xalignat 2
&\hskip -2em
T=T(s,\brbG,\overset\sssize\star\to{\bold g},\brbR,\rho),
&&T=T(t,p,s,\bold g,\hat\bold G,\bold R,\rho).\quad
\mytag{4.15}
\endxalignat
$$
Applying the inequalities \mythetag{4.14} to the functions
\mythetag{4.10} and \mythetag{4.11}, we derive
$$
c=T\,\frac{\partial s}{\partial T}>0.
$$
Here $c$ is the {\it specific heat capacity} of a medium for the case 
where the density $\rho$ and the other arguments $t$, $p$, $\bold g$, 
$\hat\bold G$, $\bold R$, $\brbG$, $\overset\sssize\star\to{\bold g}$, 
$\brbR$ in \mythetag{4.10} and \mythetag{4.11} are constants.\par
     Let's substitute the temperature expressed in the form of the
function \mythetag{4.15} into \mythetag{4.8}. As a result we get
two functions 
$$
\xalignat 2
&\hskip -2em
\varepsilon=E(s,\brbG,\overset\sssize\star\to{\bold g},\brbR,\rho),
&&\varepsilon=\varepsilon(t,p,s,\bold g,\hat\bold G,\bold R,\rho).\quad
\mytag{4.16}
\endxalignat
$$
Passing from \mythetag{2.8} and \mythetag{2.12} to the functions
\mythetag{4.16}, we perform the so-called {\it Legendre transformations}.
The functions \mythetag{4.15} are related to each other through 
\mythetag{2.9}, \mythetag{2.10}, and \mythetag{2.11}. The same is true
for the functions \mythetag{4.16}. The second function in \mythetag{4.16}
is an extended scalar field, though with the slightly different set of
arguments than in \mythetag{2.12}. From \mythetag{2.9}, \mythetag{2.10},
\mythetag{2.11}, and \mythetag{4.16} we derive
$$
\align
&\hskip -2em
\vnabla^{\alpha\beta}[2]\varepsilon=\frac{\partial \varepsilon}
{\partial g_{\alpha\beta}}
=\sum^3_{i=1}\sum^3_{j=1}\frac{\partial E}{\partial\brG_{ij}}
\,\hat S^\alpha_i\,\hat S^\beta_j,
\mytag{4.17}\\
&\hskip -2em
\vnabla^{\alpha\beta}[3]\varepsilon=\frac{\partial \varepsilon}
{\partial\hat G_{\alpha\beta}\vphantom{\vrule height 10pt}}
=\sum^3_{i=1}\sum^3_{j=1}\frac{\partial E}
{\partial\overset\sssize\star\to g_{ij}}\,\hat S^\alpha_i
\,\hat S^\beta_j,
\mytag{4.18}\\
&\hskip -2em
\vnabla^\alpha_{\!\beta}[4]\varepsilon=\frac{\partial \varepsilon}
{\partial R^\beta_\alpha\vphantom{\vrule height 11pt}}
=\sum^3_{i=1}\sum^3_{j=1}\frac{\partial E}{\partial\brR^{\,i}_j}
\,\hat S^\alpha_j\,\hat T^i_\beta.
\mytag{4.19}
\endalign
$$
Applying \mythetag{4.17}, \mythetag{4.18}, \mythetag{4.19}, and 
the formula \thetag{13.15} from \mycite{6}, we get
$$
\hskip -2em
\gathered
\nabla_{\!\gamma}\varepsilon=
-\sum^3_{m=1}\sum^3_{\alpha=1}\sum^3_{\beta=1}
\vnabla^{\alpha\beta}[2]\varepsilon
\left(\hat Z^m_{\gamma\alpha}\ g_{m\beta}
+\hat Z^m_{\gamma\beta}\ g_{\alpha m}\right)-\\
-\sum^3_{m=1}\sum^3_{\alpha=1}\sum^3_{\beta=1}
\vnabla^{\alpha\beta}[3]\varepsilon\left(\hat Z^m_{\gamma\alpha}
\ \hat G_{m\beta}+\hat Z^m_{\gamma\beta}
\ \hat G_{\alpha m}\right)+\\
+\sum^3_{n=1}\sum^3_{m=1}\sum^3_{\alpha=1}\sum^3_{\beta=1}
\vnabla^\alpha_{\!\beta}[4]\varepsilon\left(\hat Z^\beta_{\gamma m}
\ R^m_\alpha-\hat Z^m_{\gamma\alpha}\ R^\beta_m\right)\!,
\endgathered
\mytag{4.20}
$$
Like the equation \mythetag{3.1}, the equation \mythetag{4.20} can be
simplified by means of \mythetag{3.7}:
$$
\hskip -2em
\hat\nabla_{\!\gamma}\varepsilon=0.
\mytag{4.21}
$$
Note that the remark following the formula \mythetag{3.8} is valid for 
the covariant derivatives $\nabla_{\!\gamma}$ and $\hat\nabla_{\!\gamma}$
in the above formulas \mythetag{4.20} and \mythetag{4.21} as well.
\mytheorem{4.2} The specific inner thermal energy $\varepsilon$
 of a dislocated
crystalline medi\-um is an extended scalar field satisfying the 
differential equation \mythetag{4.20}, or the differential equation
\mythetag{4.21} equivalent to it.
\endproclaim
From \mythetag{4.16}, \mythetag{4.17}, \mythetag{4.18}, \mythetag{4.19}, 
and from \mythetag{4.20}, applying \mythetag{3.13}, we obtain
$$
\gathered
\frac{\partial \varepsilon}{\partial t}
+\sum^3_{i=1}v^i\ \nabla_{\!i}\,\varepsilon=
-\sum^3_{m=1}\sum^3_{\alpha=1}\sum^3_{\beta=1}
\vnabla^{\alpha\beta}[2]\varepsilon
\ \bigl(\theta^{\,m}_\alpha\ g_{m\beta}
-\nabla_{\!\alpha}v^m\ g_{m\beta}\ +\\
+\ \theta^{\,m}_\beta\ g_{\alpha m}
-\nabla_{\!\beta}v^m\ g_{\alpha m}\bigr)
-\sum^3_{m=1}\sum^3_{\alpha=1}\sum^3_{\beta=1}
\vnabla^{\alpha\beta}[3]\varepsilon
\ \bigl(\theta^{\,m}_\alpha\ \hat G_{m\beta}\ -\\
-\ \nabla_{\!\alpha}v^m\ \hat G_{m\beta}+
\theta^{\,m}_\beta\ \hat G_{\alpha m}
-\nabla_{\!\beta}v^m\ \hat G_{\alpha m}\bigr)
+\sum^3_{m=1}\sum^3_{\alpha=1}\sum^3_{\beta=1}
\vnabla^\alpha_{\!\beta}[4]\varepsilon\times\\
\vspace{2ex}
\times\bigl(\theta^\beta_m\ R^m_\alpha
-\nabla_{\!m}v^\beta\ R^m_\alpha
-\theta^{\,m}_\alpha\ R^\beta_m
+\nabla_{\!\alpha}v^m\ R^\beta_m\bigr).
\endgathered\quad
\mytag{4.22}
$$
\mytheorem{4.3} The specific inner thermal energy $\varepsilon$
of a dislocated crystalline medium is an extended scalar field 
satisfying the differential equation \mythetag{4.22}.
\endproclaim
     The equalities \mythetag{4.17}, \mythetag{4.18}, and \mythetag{4.19}
should be completed with the following one being analogous to the equality
\mythetag{4.9}:
$$
\hskip -2em
T=\frac{\partial\varepsilon}{\partial s}.
\mytag{4.23}
$$
The equality \mythetag{4.23} applies to both functions \mythetag{4.16} and
to both functions \mythetag{4.15}:
$$
\align
T(s,\brbG,\overset\sssize\star\to{\bold g},\brbR,\rho)
&=\frac{\partial\varepsilon(s,\brbG,\overset\sssize\star\to{\bold g},
\brbR,\rho)}{\partial s},\\
\vspace{1ex}
T(t,p,s,\bold g,\hat\bold G,\bold R,\rho)
&=\frac{\partial\varepsilon(t,p,s,\bold g,\hat\bold G,\bold R,\rho)}
{\partial s}=\vnabla[1]\varepsilon.
\mytag{4.24}
\endalign
$$\par
     Now let's return to the balance equation \mythetag{4.5}. 
Substituting \mythetag{4.6} and \mythetag{4.7} into \mythetag{4.5},
then taking into account other two balance equations \mythetag{4.1},
\mythetag{4.2} and the formula \mythetag{4.3}, for the function
$\varepsilon$ we derive
$$
\rho\,\partial_t\varepsilon
+\sum^3_{k=1}\rho\,v^k\,\partial_k\varepsilon
=\sum^3_{i=1}\sum^3_{k=1}\left(\nabla_{\!k}v_i\,(\sigma^{ik}
+\tilde\sigma^{ik})+\nabla_{\!k}(\nabla_{\!i}T\,
\varkappa^{ik})\right).
\quad
\mytag{4.25}
$$
(see more details in \mycite{5}). All tensor fields in the balance
equations \mythetag{4.1}, \mythetag{4.2}, \mythetag{4.5} and in the
formulas \mythetag{4.3}, \mythetag{4.4}, \mythetag{4.6}, \mythetag{4.7},
\mythetag{4.25} are treated as regular (not extended) tensor fields and
covariant derivatives $\nabla_{\!k}$ and $\nabla_{\!i}$ are understood 
as regular covariant derivatives different from those in left hand sides 
of \mythetag{4.20} and \mythetag{4.22}. The partial derivative
$\partial_t\varepsilon$ is expressed through $\partial\varepsilon/
\partial t$ in \mythetag{4.22} as follows:
$$
\hskip -2em
\gathered
\partial_t\varepsilon=\frac{\partial\varepsilon}{\partial t}
+\vnabla[1]\varepsilon\ \partial_ts+\sum^3_{\alpha=1}\sum^3_{\beta=1}
\vnabla^{\alpha\beta}[2]\varepsilon\ \partial_tg_{\alpha\beta}\ +\\
+\sum^3_{\alpha=1}\sum^3_{\beta=1}\vnabla^{\alpha\beta}[3]\varepsilon\ 
\partial_t\hat G_{\alpha\beta}+\sum^3_{\alpha=1}\sum^3_{\beta=1}
\vnabla^\alpha_{\!\beta}[4]\varepsilon\ \partial_tR^\beta_\alpha+
\vnabla[5]\varepsilon\ \partial_t\rho.
\endgathered
\mytag{4.26}
$$
Similarly, for the partial derivative $\partial_k\varepsilon$ in
\mythetag{4.25} we can derive the formula expressing it through the
covariant derivative \mythetag{4.20}:
$$
\hskip -2em
\gathered
\partial_k\varepsilon=\nabla_{\!k}\varepsilon+\vnabla[1]\varepsilon\
\partial_ks+\sum^3_{\alpha=1}\sum^3_{\beta=1}
\vnabla^{\alpha\beta}[2]\varepsilon\ \nabla_{\!k}\,g_{\alpha\beta}\ +\\
+\sum^3_{\alpha=1}\sum^3_{\beta=1}\vnabla^{\alpha\beta}[3]\varepsilon\ 
\nabla_{\!k}\hat G_{\alpha\beta}+\sum^3_{\alpha=1}\sum^3_{\beta=1}
\vnabla^\alpha_{\!\beta}[4]\varepsilon\ \nabla_{\!k}R^\beta_\alpha+
\vnabla[5]\varepsilon\ \partial_k\rho.
\endgathered
\mytag{4.27}
$$
In deriving \mythetag{4.27} we used the formula \thetag{17.9} from
\mycite{6}. Now let's use \mythetag{4.24}, \mythetag{4.26}, and 
\mythetag{4.27} in order to calculate the following expression:
$$
\gathered
\partial_t\varepsilon
+\sum^3_{k=1}v^k\,\partial_k\varepsilon
=\left(\frac{\partial\varepsilon}{\partial t}
+\shave{\sum^3_{k=1}}v^k\,\nabla_{\!k}\varepsilon\right)
+T\left(\partial_ts+\shave{\sum^3_{k=1}}v^k\,\partial_ks\right)+\\
+\sum^3_{\alpha=1}\sum^3_{\beta=1}\vnabla^{\alpha\beta}[3]\varepsilon
\left(\partial_t\hat G_{\alpha\beta}+\shave{\sum^3_{k=1}}v^k\,
\nabla_{\!k}\hat G_{\alpha\beta}\right)
+\sum^3_{\alpha=1}\sum^3_{\beta=1}\vnabla^\alpha_{\!\beta}[4]\varepsilon
\ \times\\
\times\left(\partial_tR^\beta_\alpha+\shave{\sum^3_{k=1}}v^k
\,\nabla_{\!k}R^\beta_\alpha\right)
+\vnabla[5]\varepsilon\left(\partial_t\rho+\shave{\sum^3_{k=1}}v^k
\,\partial_k\rho\right)\!.
\endgathered\quad
\mytag{4.28}
$$
The terms containing $\partial_tg_{\alpha\beta}$ and 
$\nabla_{\!k}g_{\alpha\beta}$ do vanish due to the following equalities: 
$$
\xalignat 2
&\hskip -2em
\partial_tg_{\alpha\beta}=0,
&&\nabla_{\!k}g_{\alpha\beta}=0.
\mytag{4.29}
\endxalignat 
$$
The first equality \mythetag{4.29} is obvious: the metric tensor $\bold g$
is a geometric equipment of the real space $\Bbb E$, its components are
constants. The second equality \mythetag{4.29} expresses the concordance
condition for the metric and connection (see \mycite{8} and \mycite{9}).
\par
     Let's apply the formula \mythetag{4.22} and the first balance 
equation \mythetag{4.1} in order to transform the equality \mythetag{4.28}.
As a result we write it as follows:
$$
\gathered
\partial_t\varepsilon
+\sum^3_{k=1}v^k\,\partial_k\varepsilon
=T\left(\partial_ts+\shave{\sum^3_{k=1}}v^k\,\partial_ks\right)
+\sum^3_{\alpha=1}\sum^3_{\beta=1}\vnabla^{\alpha\beta}[2]\varepsilon
\ \times\\
\times\left(\nabla_{\!\alpha}v_\beta+\nabla_{\!\beta}\,v_\alpha
-\shave{\sum^3_{m=1}}\bigl(\theta^{\,m}_\alpha\ g_{m\beta}
+\theta^{\,m}_\beta\ g_{\alpha m}\bigr)\!\right)
+\sum^3_{\alpha=1}\sum^3_{\beta=1}\vnabla^{\alpha\beta}[3]
\varepsilon\ \times\\
\times\left(\partial_t\hat G_{\alpha\beta}+\shave{\sum^3_{k=1}}v^k\,
\nabla_{\!k}\hat G_{\alpha\beta}+\shave{\sum^3_{m=1}}
\nabla_{\!\alpha}v^m\ \hat G_{m\beta}+\shave{\sum^3_{m=1}}
\nabla_{\!\beta}\,v^m\ \hat G_{\alpha m}\ -\right.\\
\left.-\shave{\sum^3_{m=1}}\theta^{\,m}_\alpha\ \hat G_{m\beta}
-\shave{\sum^3_{m=1}}\theta^{\,m}_\beta\ \hat G_{\alpha m}\right)
+\sum^3_{\alpha=1}\sum^3_{\beta=1}\vnabla^\alpha_{\!\beta}[4]\varepsilon
\left(\partial_tR^\beta_\alpha\ +\vphantom{\shave{\sum^3_{k=1}}}\right.\\
+\shave{\sum^3_{k=1}}v^k\,\nabla_{\!k}R^\beta_\alpha
-\shave{\sum^3_{m=1}}\nabla_{\!m}v^\beta\ R^m_\alpha
+\shave{\sum^3_{m=1}}\nabla_{\!\alpha}v^m\ R^\beta_m
+\shave{\sum^3_{m=1}}\theta^\beta_m\ R^m_\alpha\ -\\
\left.-\shave{\sum^3_{m=1}}\theta^{\,m}_\alpha\ R^\beta_m\right)
+\shave{\sum^3_{k=1}}\vnabla[5]\varepsilon\ \rho\ \nabla_{\!k}v^k.
\endgathered\quad
\mytag{4.30}
$$
Now it is worth to remember the differential equation for the elastic 
deformation tensor $\hat\bold G$. It was suggested empirically for
plastic media in \mycite{5}. Later it was derived for dislocated 
crystalline media in \mycite{1} (see formula \thetag{3.8} over there):
$$
\hskip -2em
\gathered
\frac{\partial\hat G_{kq}}{\partial t}+\sum^3_{r=1}v^r\,
\nabla_{\!r}\hat G_{kq}=-\sum^3_{r=1}\nabla_{\!k}v^r\,
\hat G_{rq}-\sum^3_{r=1}\hat G_{kr}\,\nabla_{\!q}v^r+\\
+\sum^3_{r=1}\theta^{\,r}_k\,\hat G_{rq}+\sum^3_{r=1}
\hat G_{kr}\,\theta^{\,r}_q.
\endgathered
\mytag{4.31}
$$
Comparing \mythetag{4.31} and \mythetag{4.30}, we find that the term
with $\vnabla^{\alpha\beta}[3]\varepsilon$ in \mythetag{4.30} vanishes
due to the equation \mythetag{4.31}. As a result we get
$$
\gathered
\partial_t\varepsilon
+\sum^3_{k=1}v^k\,\partial_k\varepsilon
=T\left(\partial_ts+\shave{\sum^3_{k=1}}v^k\,\partial_ks\right)
+\sum^3_{\alpha=1}\sum^3_{\beta=1}\vnabla^{\alpha\beta}[2]\varepsilon
\ \times\\
\times\left(\nabla_{\!\alpha}v_\beta+\nabla_{\!\beta}\,v_\alpha
-\shave{\sum^3_{m=1}}\bigl(\theta^{\,m}_\alpha\ g_{m\beta}
+\theta^{\,m}_\beta\ g_{\alpha m}\bigr)\!\right)
+\sum^3_{\alpha=1}\sum^3_{\beta=1}\vnabla^\alpha_{\!\beta}[4]\varepsilon
\ \times\\
\times\left(\partial_tR^\beta_\alpha
+\shave{\sum^3_{k=1}}v^k\,\nabla_{\!k}R^\beta_\alpha
-\shave{\sum^3_{m=1}}\nabla_{\!m}v^\beta\ R^m_\alpha
+\shave{\sum^3_{m=1}}\nabla_{\!\alpha}v^m\ R^\beta_m\ +\right.\\
\left.+\shave{\sum^3_{m=1}}\theta^\beta_m\ R^m_\alpha\ 
-\shave{\sum^3_{m=1}}\theta^{\,m}_\alpha\ R^\beta_m\right)
+\sum^3_{\alpha=1}\sum^3_{\beta=1}\vnabla[5]\varepsilon\ \rho\ 
g^{\alpha\beta}\ \nabla_{\!\alpha}v_\beta.
\endgathered\quad
\mytag{4.32}
$$
In the next step we should recall the differential equation that 
determines the time evolution of the tensor $\bold R$ (see \thetag{4.11} 
in \mycite{3} or \thetag{4.1} in \mycite{4}):
$$
\gathered
\frac{\partial R^\beta_\alpha}{\partial t}
=\sum^3_{m=1}J^{\,\beta}_m\,R^m_\alpha
+\sum^3_{m=1}\nabla_{\!m}v^\beta\,R^m_\alpha
+\sum^3_{m=1}\sum^3_{p=1}v^p\,\hat Z^\beta_{mp}\,R^m_\alpha\,-\\
-\sum^3_{m=1}\sum^3_{r=1}\sum^3_{s=1}g_{\alpha m}\,\omega^{mrs}
\,\nabla_{\!r}J^{\,\beta}_{\,s}-\sum^3_{m=1}\sum^3_{r=1}\sum^3_{s=1}
\sum^3_{p=1}g_{\alpha m}\,\omega^{mrs}\,\hat Z^\beta_{rp}
\,J^{\,p}_{\,s}.
\endgathered\quad
\mytag{4.33}
$$
Let's apply the relationship \mythetag{3.12} to the term $J^{\,\beta}_m$
in the first sum of \mythetag{4.33}:
$$
\gathered
\frac{\partial R^\beta_\alpha}{\partial t}
=-\sum^3_{m=1}\theta^{\,\beta}_m\,R^m_\alpha
+\sum^3_{m=1}\nabla_{\!m}v^\beta\,R^m_\alpha
+\sum^3_{m=1}\sum^3_{p=1}v^p\,\hat Z^\beta_{p\,m}\,R^m_\alpha\,-\\
-\sum^3_{m=1}\sum^3_{r=1}\sum^3_{s=1}g_{\alpha m}\,\omega^{mrs}
\,\nabla_{\!r}J^{\,\beta}_{\,s}-\sum^3_{m=1}\sum^3_{r=1}\sum^3_{s=1}
\sum^3_{p=1}g_{\alpha m}\,\omega^{mrs}\,\hat Z^\beta_{rp}
\,J^{\,p}_{\,s}.
\endgathered\quad
\mytag{4.34}
$$
Then let's recall that the formula \mythetag{3.12} can be written in a
different form:
$$
\hskip -2em
J^{\,p}_s=-\theta^{\,p}_s+\sum^3_{\tau=1}\sum^3_{\gamma=1}\sum^3_{n=1}
\omega_{\tau\gamma s}\ v^\gamma\ g^{\tau n}\,R^p_n
\mytag{4.35}
$$
(see formula \thetag{2.13} in \mycite{3}). Before substituting
\mythetag{4.35} into the equation \mythetag{4.34} let's recall the
following purely algebraic identity (it was already used in \mycite{2}
and \mycite{3}):
$$
\hskip -2em
\sum^3_{s=1}\omega_{\tau\gamma s}\,\omega^{mrs}=
\delta^m_\tau\,\delta^r_\gamma-\delta^m_\gamma\,\delta^r_\tau.
\mytag{4.36}
$$
Here $\delta^m_\tau$, $\delta^r_\gamma$, $\delta^m_\gamma$, and
$\delta^r_\tau$ are Kronecker $\delta$-symbols. Now substituting
\mythetag{4.35} into the last term of \mythetag{4.34} and taking 
into account \mythetag{4.36}, we derive
$$
\gather
\sum^3_{m=1}\sum^3_{r=1}\sum^3_{s=1}
\sum^3_{p=1}g_{\alpha m}\,\omega^{mrs}\,\hat Z^\beta_{rp}
\,J^{\,p}_{\,s}=-\sum^3_{m=1}\sum^3_{r=1}\sum^3_{s=1}
\sum^3_{p=1}g_{\alpha m}\,\omega^{mrs}\,\hat Z^\beta_{rp}
\ \theta^{\,p}_s\ +\\
+\sum^3_{m=1}\sum^3_{r=1}\sum^3_{s=1}\sum^3_{p=1}\sum^3_{\tau=1}
\sum^3_{\gamma=1}\sum^3_{n=1}g_{\alpha m}\,\omega^{mrs}
\,\hat Z^\beta_{rp}\,\omega_{\tau\gamma s}\ v^\gamma\ g^{\tau n}
\,R^p_n=\\
=-\sum^3_{m=1}\sum^3_{r=1}\sum^3_{s=1}
\sum^3_{p=1}g_{\alpha m}\,\omega^{mrs}\,\hat Z^\beta_{rp}
\ \theta^{\,p}_s\,+\sum^3_{m=1}\sum^3_{r=1}\sum^3_{p=1}
\sum^3_{\tau=1}\sum^3_{\gamma=1}\sum^3_{n=1}g_{\alpha m}\,\times\\
\times\left(\delta^m_\tau\,\delta^r_\gamma-\delta^m_\gamma
\,\delta^r_\tau\right)\hat Z^\beta_{rp}\ v^\gamma\ g^{\tau n}
\,R^p_n=-\sum^3_{m=1}\sum^3_{r=1}\sum^3_{s=1}
\sum^3_{p=1}g_{\alpha m}\,\omega^{mrs}\,\hat Z^\beta_{rp}\ \times\\
\times\ \theta^{\,p}_s\,+
\sum^3_{r=1}\sum^3_{p=1}v^r\,\hat Z^\beta_{rp}\ R^p_\alpha
-\sum^3_{r=1}\sum^3_{p=1}\sum^3_{n=1}v_\alpha\,
\hat Z^\beta_{rp}\ g^{rn}\ R^p_n.
\endgather
$$
Applying this result to the equation \mythetag{4.34}, we can write
this equation as follows:
$$
\gathered
\frac{\partial R^\beta_\alpha}{\partial t}
=-\sum^3_{m=1}\theta^{\,\beta}_m\,R^m_\alpha
+\sum^3_{m=1}\nabla_{\!m}v^\beta\,R^m_\alpha
+\sum^3_{r=1}\sum^3_{p=1}\sum^3_{n=1}v_\alpha\,
\hat Z^\beta_{rp}\ g^{rn}\ R^p_n\ -\\
-\sum^3_{m=1}\sum^3_{r=1}\sum^3_{s=1}g_{\alpha m}\,\omega^{mrs}
\,\nabla_{\!r}J^{\,\beta}_{\,s}+\sum^3_{m=1}\sum^3_{r=1}\sum^3_{s=1}
\sum^3_{p=1}g_{\alpha m}\,\omega^{mrs}\,\hat Z^\beta_{rp}
\ \theta^{\,p}_s.
\endgathered\quad
\mytag{4.37}
$$
Now let's apply the operator $\nabla_{\!r}$ to the equality
\mythetag{4.35}. As a result we obtain
$$
\hskip -2em
\gathered
\nabla_{\!r}J^{\,\beta}_s=-\nabla_{\!r}\theta^{\,\beta}_s
+\sum^3_{\tau=1}\sum^3_{\gamma=1}\sum^3_{n=1}
\omega_{\tau\gamma s}\ \nabla_{\!r}v^\gamma\ g^{\tau n}
\,R^\beta_n\ +\\
+\sum^3_{\tau=1}\sum^3_{\gamma=1}\sum^3_{n=1}
\omega_{\tau\gamma s}\ v^\gamma\ g^{\tau n}
\,\nabla_{\!r}R^\beta_n.
\endgathered
\mytag{4.38}
$$
Taking into account \mythetag{4.38} and \mythetag{4.36}, we can perform 
the following calculations:
$$
\allowdisplaybreaks
\gather
\sum^3_{m=1}\sum^3_{r=1}\sum^3_{s=1}g_{\alpha m}\ \omega^{mrs}
\,\nabla_{\!r}J^{\,\beta}_{\,s}=
-\sum^3_{m=1}\sum^3_{r=1}\sum^3_{s=1}g_{\alpha m}\ \omega^{mrs}
\,\nabla_{\!r}\theta^{\,\beta}_{\,s}\ +\\
+\sum^3_{m=1}\sum^3_{r=1}\sum^3_{s=1}\sum^3_{\tau=1}\sum^3_{\gamma=1}
\sum^3_{n=1}g_{\alpha m}\ \omega^{mrs}\ \omega_{\tau\gamma s}\ 
\nabla_{\!r}v^\gamma\ g^{\tau n}\ R^\beta_n\ +\\
+\sum^3_{m=1}\sum^3_{r=1}\sum^3_{s=1}\sum^3_{\tau=1}\sum^3_{\gamma=1}
\sum^3_{n=1}g_{\alpha m}\ \omega^{mrs}\ \omega_{\tau\gamma s}\ 
v^\gamma\ g^{\tau n}\ \nabla_{\!r}R^\beta_n=
-\sum^3_{m=1}\sum^3_{r=1}\sum^3_{s=1}g_{\alpha m}\,\times\\
\times\ \omega^{mrs}\ \nabla_{\!r}\theta^{\,\beta}_{\,s}
+\sum^3_{m=1}\sum^3_{r=1}\sum^3_{\tau=1}\sum^3_{\gamma=1}
\sum^3_{n=1}g_{\alpha m}\left(\delta^m_\tau\,\delta^r_\gamma
-\delta^m_\gamma\,\delta^r_\tau\right)\nabla_{\!r}v^\gamma
\ g^{\tau n}\ R^\beta_n\ +\\
+\sum^3_{m=1}\sum^3_{r=1}\sum^3_{\tau=1}\sum^3_{\gamma=1}
\sum^3_{n=1}g_{\alpha m}\left(\delta^m_\tau\,\delta^r_\gamma
-\delta^m_\gamma\,\delta^r_\tau\right)v^\gamma
\ g^{\tau n}\ \nabla_{\!r}R^\beta_n=\\
=-\sum^3_{m=1}\sum^3_{r=1}\sum^3_{s=1}g_{\alpha m}
\ \omega^{mrs}\ \nabla_{\!r}\theta^{\,\beta}_{\,s}
+\sum^3_{r=1}\nabla_{\!r}v^r\ R^\beta_\alpha
-\sum^3_{r=1}\sum^3_{n=1}\nabla_{\!r}v_\alpha
\ g^{rn}\ R^\beta_n\ +\\
+\sum^3_{r=1}v^r\ \nabla_{\!r}R^\beta_\alpha
-\sum^3_{r=1}\sum^3_{n=1}v_\alpha\ g^{rn}\ \nabla_{\!r}R^\beta_n.
\endgather
$$
The result of these calculations is used in order to transform the
equation \mythetag{4.37}:
$$
\gathered
\frac{\partial R^\beta_\alpha}{\partial t}
=-\sum^3_{m=1}\theta^{\,\beta}_m\,R^m_\alpha
+\sum^3_{m=1}\nabla_{\!m}v^\beta\,R^m_\alpha
+\sum^3_{r=1}\sum^3_{p=1}\sum^3_{n=1}v_\alpha\,
\hat Z^\beta_{rp}\ g^{rn}\,R^p_n\ -\\
-\sum^3_{r=1}\left(\nabla_{\!r}v^r\ R^\beta_\alpha
+v^r\ \nabla_{\!r}R^\beta_\alpha\,\right)
+\sum^3_{r=1}\sum^3_{n=1}g^{rn}\left( 
\nabla_{\!r}v_\alpha\ R^\beta_n
+v_\alpha\ \nabla_{\!r}R^\beta_n\,\right)\ +\\
+\sum^3_{m=1}\sum^3_{r=1}\sum^3_{s=1}g_{\alpha m}\,\omega^{mrs}
\ \nabla_{\!r}\theta^{\,\beta}_s
+\sum^3_{m=1}\sum^3_{r=1}\sum^3_{s=1}
\sum^3_{p=1}g_{\alpha m}\,\omega^{mrs}\,\hat Z^\beta_{rp}
\ \theta^{\,p}_s.
\endgathered\quad
\mytag{4.39}
$$
The following identity is known as the {\it zero divergency condition}
(see \thetag{4.8} in \mycite{3}):
$$
\hskip -2em
\sum^3_{r=1}\sum^3_{r=1}g^{rn}\,\nabla_{\!r}R^\beta_n
+\sum^3_{p=1}\sum^3_{r=1}\sum^3_{n=1}\hat Z^\beta_{rp}
\ g^{rn}\,R^p_n=0.
\mytag{4.40}
$$
Applying \mythetag{4.40} to \mythetag{4.39}, we find that two terms
in \mythetag{4.39} are canceled:
$$
\gathered
\frac{\partial R^\beta_\alpha}{\partial t}
=\sum^3_{m=1}\nabla_{\!m}v^\beta\,R^m_\alpha
-\sum^3_{r=1}\nabla_{\!r}v^r\ R^\beta_\alpha
-\sum^3_{r=1}v^r\ \nabla_{\!r}R^\beta_\alpha\ +\\
+\sum^3_{r=1}\sum^3_{n=1}g^{rn}\ \nabla_{\!r}v_\alpha\ R^\beta_n
+\sum^3_{m=1}\sum^3_{r=1}\sum^3_{s=1}g_{\alpha m}\,\omega^{mrs}
\ \nabla_{\!r}\theta^{\,\beta}_s\ +\\
+\sum^3_{m=1}\sum^3_{r=1}\sum^3_{s=1}
\sum^3_{p=1}g_{\alpha m}\,\omega^{mrs}\,\hat Z^\beta_{rp}
\ \theta^{\,p}_s-\sum^3_{m=1}\theta^{\,\beta}_m\,R^m_\alpha.
\endgathered\quad
\mytag{4.41}
$$\par 
     Having completed the transformations of the differential 
equation \mythetag{4.33}, now we return to the formula \mythetag{4.32}.
Substituting \mythetag{4.41} into \mythetag{4.32}, we obtain
$$
\allowdisplaybreaks
\gather
\partial_t\varepsilon
+\sum^3_{k=1}v^k\,\partial_k\varepsilon
=T\left(\partial_ts+\shave{\sum^3_{k=1}}v^k\,\partial_ks\right)
+\sum^3_{\alpha=1}\sum^3_{\beta=1}\vnabla^{\alpha\beta}[2]\varepsilon
\ \times\\
\times\left(\nabla_{\!\alpha}v_\beta+\nabla_{\!\beta}\,v_\alpha
-\shave{\sum^3_{m=1}}\bigl(\theta^{\,m}_\alpha\ g_{m\beta}
+\theta^{\,m}_\beta\ g_{\alpha m}\bigr)\!\right)
+\sum^3_{\alpha=1}\sum^3_{\beta=1}\vnabla^\alpha_{\!\beta}[4]\varepsilon
\ \times\\
\times\left(\,\shave{\sum^3_{m=1}}\nabla_{\!\alpha}v^m\ R^\beta_m
-\shave{\sum^3_{r=1}}\nabla_{\!r}v^r\ R^\beta_\alpha
+\shave{\sum^3_{r=1}}\shave{\sum^3_{n=1}}g^{rn}\ \nabla_{\!r}v_\alpha
\ R^\beta_n\ +\right.\\
+\sum^3_{m=1}\sum^3_{r=1}\sum^3_{s=1}g_{\alpha m}\,\omega^{mrs}
\ \nabla_{\!r}\theta^{\,\beta}_s
+\sum^3_{m=1}\sum^3_{r=1}\sum^3_{s=1}
\sum^3_{p=1}g_{\alpha m}\,\omega^{mrs}\,\hat Z^\beta_{rp}
\ \theta^{\,p}_s\ -\\
\left.-\shave{\sum^3_{m=1}}\theta^{\,m}_\alpha\ R^\beta_m\right)
+\sum^3_{\alpha=1}\sum^3_{\beta=1}\vnabla[5]\varepsilon\ \rho\ 
g^{\alpha\beta}\ \nabla_{\!\alpha}v_\beta.
\endgather
$$
Let's perform some rearrangement of terms in the above equality.
We write it as
$$
\gathered
\partial_t\varepsilon
+\sum^3_{k=1}v^k\,\partial_k\varepsilon
=T\left(\partial_ts+\shave{\sum^3_{k=1}}v^k\,\partial_ks\right)
+\sum^3_{i=1}\sum^3_{j=1}\sum^3_{k=1}\ T\ \times\\
\times\ \frac{\nabla_{\!k}\left(P^{k\kern 0.5pt ij}\ \theta_{ij}\right)}
{\rho}+\sum^3_{i=1}\sum^3_{j=1}\frac{\sigma^{ij}\ \nabla_{\!i}\,v_j}
{\rho}-\sum^3_{i=1}\sum^3_{j=1}\frac{\goth S^{ij}\ \theta_{ij}}{\rho}.
\endgathered
\mytag{4.42}
$$
For this purpose we introduce some auxiliary notations. We denote
$$
\hskip -2em
\theta_{ij}=\sum^3_{k=1}g_{ik}\ \theta^{\,k}_j.
\mytag{4.43}
$$
It is easy to see that the formula \mythetag{4.42} expresses 
the standard index lowering procedure. The other notations 
are less trivial. We denote
$$
\hskip -2em
P^{k\kern 0.5pt ij}=\sum^3_{\alpha=1}\sum^3_{\beta=1}\sum^3_{m=1}
\frac{\rho\ g^{\beta i}\ g_{\alpha m}\ \omega^{mkj}}{T}
\ \vnabla^\alpha_{\!\beta}[4]\varepsilon.
\mytag{4.44}
$$
The quantities \mythetag{4.43} define a tensor field of the type $(3,0)$.
Then we introduce the symmetric tensor field $\boldsymbol\sigma$ of the 
type $(2,0)$:
$$
\gather
\hskip -2em
\sigma^{ij}=\rho\ \bigl(\vnabla^{ij}[2]\varepsilon
+\vnabla^{j\,i}[2]\varepsilon\bigr)\ +\\
\vspace{1ex}
\hskip -2em
+\sum^3_{\alpha=1}\sum^3_{\beta=1}\rho\ \bigl(\vnabla^j_{\!\beta}[4]
\varepsilon\ R^\beta_\alpha\ g^{\alpha i}
+\vnabla^{\kern 1pt i}_{\!\beta}[4]\varepsilon\ R^\beta_\alpha
\ g^{\alpha j}\bigr)\ -
\mytag{4.45}\\
\hskip -2em
-\sum^3_{\alpha=1}\sum^3_{\beta=1}\rho\ \vnabla^\alpha_{\!\beta}
[4]\varepsilon\ R^\beta_\alpha\ g^{ij}
+\rho^2\ \vnabla[5]\varepsilon\ g^{ij}.
\endgather
$$
And finally, we introduce another tensor field of the type $(2,0)$
with the components
$$
\pagebreak
\gathered
\goth S^{ij}=\rho\ \bigl(\vnabla^{ij}[2]\varepsilon
+\vnabla^{j\,i}[2]\varepsilon\bigr)
+\sum^3_{\alpha=1}\sum^3_{\beta=1}\rho\ \vnabla^j_{\!\beta}[4]
\varepsilon\ R^\beta_\alpha\ g^{\alpha i}\ -\\
-\sum^3_{\alpha=1}\sum^3_{\beta=1}\sum^3_{m=1}\sum^3_{r=1}
\sum^3_{p=1}\rho\ \vnabla^\alpha_{\!\beta}[4]\varepsilon\ g_{\alpha m}
\ \omega^{mrj}\ \hat Z^\beta_{rp}\ g^{p\,i}+\sum^3_{k=1}T
\ \nabla_{\!k}P^{k\kern 0.5pt ij}.
\endgathered\quad
\mytag{4.46}
$$
Note that \mythetag{4.45} is not simply a notation. The same symbol 
$\sigma$ is used in \mythetag{4.25} and in \mythetag{4.3}, where it
represents the components of the regular stress tensor.
\mytheorem{4.4} In the case of frozen dislocations the components of 
the regular stress tensor $\boldsymbol\sigma$ are given by the formula
\mythetag{4.45}.
\endproclaim
\demo{Proof} Let's remember that the case of frozen dislocations is that
very case where the dislocation lines move together with the medium like 
water-plants frozen into the ice (see \mycite{2}). In this case we have
purely elastic response of the medium:
$$
\hskip -2em
\theta^{\,k}_j=0
\mytag{4.47}
$$
(see theorem~2.2 in \mycite{3}). From \mythetag{4.43} and \mythetag{4.47}
we derive
$$
\hskip -2em
\theta_{ij}=0.
\mytag{4.48}
$$
Let's substitute \mythetag{4.48} into \mythetag{4.42} and replace 
$\sigma^{ij}$ by $\hat\sigma^{ij}$ in this formula:
$$
\partial_t\varepsilon
+\sum^3_{k=1}v^k\,\partial_k\varepsilon
=T\left(\partial_ts+\shave{\sum^3_{k=1}}v^k\,\partial_ks\right)
+\sum^3_{i=1}\sum^3_{j=1}\frac{\hat\sigma^{ij}
\ \nabla_{\!i}\,v_j}{\rho}.
\mytag{4.49}
$$
Then we substitute \mythetag{4.49} into \mythetag{4.25}. Taking into 
account \mythetag{4.3}, \mythetag{4.4}, and the first balance equation
\mythetag{4.1}, from we \mythetag{4.25} derive 
$$
\gathered
\frac{\partial(\,\rho\,s)}{\partial t}+\sum^3_{k=1}\nabla_{\!k}\!
\left(\rho\,s\,v^k-\shave{\sum^3_{i=1}}\frac{\nabla_{\!i}T\,
\varkappa^{ik}}{T}\right)=\sum^3_{i=1}\sum^3_{j=1}\frac{(\sigma^{ij}
-\hat\sigma^{ij})\,v_{ij}}{T}\,+\\
+\sum^3_{i=1}\sum^3_{k=1}\sum^3_{j=1}\sum^3_{q=1}\frac{v_{ik}\,
\eta^{ikjq}\,v_{jq}}{T}+\sum^3_{i=1}\sum^3_{k=1}\frac{\nabla_{\!i}T
\,\varkappa^{ik}\,\nabla_{\!k}T}{T^2},
\endgathered
\quad
\mytag{4.50}
$$
where $v_{ij}$ are determined by the following formula:
$$
\hskip -2em
v_{ij}=\frac{\nabla_{\!i}v_j+\nabla_{\!j}v_i}{2}.
\mytag{4.51}
$$
The equation \mythetag{4.50} has a quite transparent interpretation.
This is the entropy balance equation. Each term in its right hand side
corresponds to some definite mechanism of entropy production. Last two 
terms correspond to viscosity and heat conductivity phenomena respectively.
In the case of frozen dislocations, i\.\,e\. in the case of purely
elastic medium response, we have no additional entropy production
mechanisms. Hence, the first term in the right hand side of \mythetag{4.50}
should vanish:
$$
\hskip -2em
\sigma^{ij}-\hat\sigma^{ij}=0.
\mytag{4.52}
$$
The equality \mythetag{4.52} means that $\hat\sigma^{ij}$ coincides
with $\sigma^{ij}$, i\.\,e\. \mythetag{4.45} is a true expression for
the components of the regular stress tensor in the case of frozen 
dislocations. The theorem is proved.
\qed\enddemo
\myconjecture{4.1} The components of the regular stress tensor
$\boldsymbol\sigma$ for a dislocated crystalline medium are always 
given by the formula \pagebreak\mythetag{4.45}.
\endproclaim
     Assuming that the conjecture~\mytheconjecture{4.1} is valid, 
we substitute \mythetag{4.42} into the equality \mythetag{4.25}. 
As a result from \mythetag{4.25} we derive the entropy balance
equation
$$
\gathered
\frac{\partial(\,\rho\,s)}{\partial t}+\sum^3_{k=1}\nabla_{\!k}\!
\left(\rho\,s\,v^k-\shave{\sum^3_{i=1}}\frac{\nabla_{\!i}T\,
\varkappa^{ik}}{T}+\shave{\sum^3_{i=1}}\shave{\sum^3_{j=1}}
P^{k\kern 0.5pt ij}\ \theta_{ij}\right)=\\
=\sum^3_{i=1}\sum^3_{j=1}\frac{\goth S^{ij}\,\theta_{ij}}{T}
+\sum^3_{i=1}\sum^3_{k=1}\sum^3_{j=1}\sum^3_{q=1}\frac{v_{ik}\,
\eta^{ikjq}\,v_{jq}}{T}+\sum^3_{i=1}\sum^3_{k=1}\frac{\nabla_{\!i}T
\,\varkappa^{ik}\,\nabla_{\!k}T}{T^2},
\endgathered\quad
\mytag{4.53}
$$
where $v_{ik}$ and $v_{jq}$ are determined by the formula \mythetag{4.51}.
\head
5. Some conclusions.
\endhead
    The equation \mythetag{4.53} is a basic equation for 
understanding the thermodynamics of plasticity in crystals. 
It is similar to the equation \thetag{10.13} in \mycite{5}. 
However, there are some visible differences. In the left hand 
side of \mythetag{4.53} we have the additional term produced 
by the tensor $\bold P$ with the components \mythetag{4.44}. 
It describes the entropy carried by moving dislocations. As 
for the first term in the right hand side of \mythetag{4.53}, 
it is also different from such a term in \mycite{5} since
$\sigma^{ij}\neq\goth S^{ij}$ (compare \mythetag{4.45} and 
\mythetag{4.46} above). The entropy growth  condition leads 
to the inequality
$$
\sum^3_{i=1}\sum^3_{j=1}\frac{\goth S^{ij}\,\theta_{ij}}{T}
\geqslant 0.
$$
It is similar to \thetag{10.14} in \mycite{5}. Note that in 
the present theory we have no restrictions like $\det\hat
\bold G=\det\bold G$ and $\det\check\bold G=1$. We got rid
of them by including the density $\rho$ as an explicit argument 
in \mythetag{2.8} and \mythetag{2.12}. As a result we have
no restriction for the trace of the matrix $\theta^{\,i}_j$.
The symmetry condition for $\theta_{ij}$ is also absent.

\enddocument
\end